\documentstyle[11pt,newpasp]{article}

\begin{document}

\title{An ISOCAM Mid-IR Survey through Gravitationally Lensing Galaxy Clusters}

\author{A. Biviano}
\affil{Osservatorio Astronomico di Trieste, via G.B. Tiepolo 11, I-34134 Trieste,
Italy}
\author{L. Metcalfe, B. Altieri, K. Leech, B. Schulz, M.F. Kessler}
\affil{ESA Astrophysics Division, Villafranca del Castillo, PO Box 50727,
E-28080 Madrid, Spain}
\author{B. McBreen, M. Delaney}
\affil{Physics Dept., University College Dublin, Stillorgan Rd., Dublin 4,
Ireland}
\author{J.-P. Kneib, G. Soucail}
\affil{Observatoire Midi-Pyr\'ene\'es, 14 Av. E. Belin, F-31400 Toulouse,
France}
\author{K. Okumura}
\affil{Institut d'Astrophysique Spatiale, Bat. 121, Universit\'e Paris-Sud,
F-91405  Orsay, France}
\author{D. Elbaz}
\affil{DSM/DAPNIA/Sap, CEA-Saclay, 91191, Gyf-sur-Yvette Cedex,
France}
\author{H. Aussel}
\affil{Osservatorio Astronomico di Padova, vicolo dell'Osservatorio 5, 
I-35122 Padova, Italy}

\begin{abstract}
We present imaging results and source counts from a deep ISOCAM
cosmological survey at 15$\mu$m, through gravitationally lensing
galaxy clusters. We take advantage of the cluster gravitational
amplification to increase the sensitivity of our survey. We detect a
large number of luminous mid-IR sources behind the cluster lenses,
down to very faint fluxes, which would have been unreachable without
the gravitational lensing effect. These source counts, corrected for
lensing distortion effects and incompleteness, are in excess of the
predictions of no-evolution models that fit local IRAS counts. By
integrating the 15 $\mu$m source counts from our counts limit, 30
$\mu$Jy, to 50 mJy we estimate the resolved mid-IR background
radiation intensity.
\end{abstract}


\section{Introduction}
The high sensitivity of the CAMera on board the ISO\footnote{ISO is an
ESA project with instruments funded by ESA Member States (especially
the PI countries: France, Germany, the Netherlands and the United
Kingdom) and with the participation of ISAS and NASA} satellite
(Cesarsky et al. 1996) has allowed detection of distant $z \sim 1$
galaxies at mid-infrared (MIR hereafter) wavelengths. These
detections are crucial for understanding galaxy evolution, since
theoretical models predict that dust obscuration can be quite an
important effect in high redshift galaxies (e.g. Guiderdoni et
al. 1997) and the dust-processed stellar radiation is re-emitted in
the IR.

The evidence emerging from various IR number count surveys
(e.g. Aussel et al. 1999, Elbaz et al. 1999, Flores et al. 1999)
indicates that strong IR emitters at $z \sim 1$ are an order of
magnitude more numerous than the extrapolation from local IRAS counts
indicate, when assuming no evolution.

In this paper we report key results of a very deep ISOCAM survey we
have conducted in three cluster fields. We took advantage of the
gravitational lensing amplification by the cluster potential wells to
detect the intrinsically faintest sources ever detected at 15 $\mu$m.
A description of our survey and results can be found in Altieri et
al. (1999) and Metcalfe et al. (1999), and even more refined analysis
is ongoing.

\section{Observations}

We observed the fields of three well-studied gravitationally lensing
clusters (Abell 370, Abell 2218, and Abell 2390) during $\sim 40$
hours of ISO guaranteed time. The three fields were imaged in two
(wide) ISOCAM filters, LW2 and LW3, centered at about 7 and 15 $\mu$m,
respectively. We used a pixel-field-of-view of 3$^{\prime\prime}$ and
micro-rastering with 7$^{\prime\prime}$ steps. This ensured good
astrometric/positional results (essential for cross-correlating these
mid-IR images with optical images, and for future observational
follow-ups).  The total area covered by our survey is $\sim 56$
arcmin$^2$.


For what concerns data reduction, source detection and photometry,
error and completeness estimates, we refer the reader to Altieri et
al. (1998, 1999). For reasons of space we here present only results
obtained with the LW3 filter, the longest wavelength ISOCAM filter.

\section{Results}

Gravitational lensing has two (equally important) effects: it
suppresses confusion because of surface area dilation, and it
amplifies the flux from background sources (see, e.g., McBreen \&
Metcalfe 1987, Paczynski 1987).  Clearly, detailed modelling of the
lens is needed in order to recover the intrinsic fluxes of the lensed
galaxies, and their space density.  For this reason, we accurately
chose our targets among the best studied clusters 
where gravitational lensing has been detected.

We achieve apparent 5~$\sigma$ sensitivities (i.e. ignoring the effect
of lensing) of 67 $\mu$Jy at 15 $\mu$m in the deepest field
(A2390),  and 80~\% completeness levels (before accounting
for lensing) of 100, 250 and 500 $\mu$Jy in the fields of A2390, A2218
and A370, respectively. In order to correct
for lensing, we use detailed models of the three cluster lenses (see
Kneib et al. 1996, B\'ezecourt et al. 1999). The highest lensing gains
are found to be $\sim 10$. The {\em intrinsically} faintest detected
source (i.e. after correcting for lensing amplification) is an 18
$\mu$Jy source, lensed to an apparent 80 $\mu$Jy source (a 6~$\sigma$
detection).

\begin{figure*}
\includegraphics{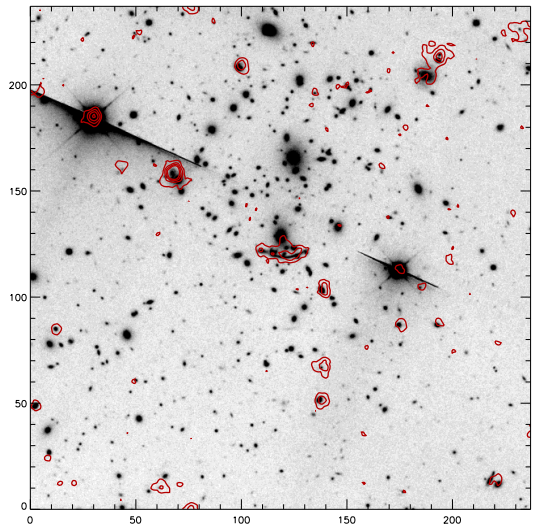}
\vspace{7cm}
\caption{ISOCAM 15 $\mu$m contours over a deep CFHT I-band image of
A370. Note the ISOCAM detection of the giant arc near the center of
the field. Units on both axes are in arcsec. From Metcalfe et al.}
\end{figure*}

\begin{figure*}
\includegraphics{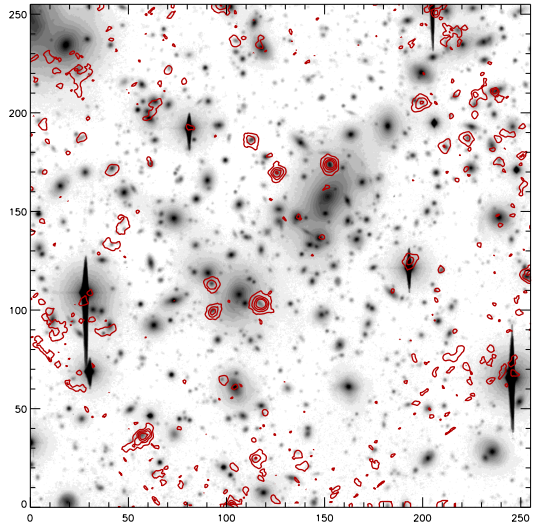}
\vspace{7cm}
\caption{ISOCAM 15 $\mu$m contours over a deep Palomar 5m I-band image
of A2218. Note the ISOCAM detection of the giant arc at (117,105).
 Units on both axes are in arcsec. From Metcalfe et al. (1999).}
\end{figure*}

In total, 71 MIR sources are detected at 5~$\sigma$ over the three
cluster fields. Most of them can be identified with (often rather
faint) visual counterparts. While they rarely correspond to the
optical arc(let)s, there are cases of impressive correlations between
optical and MIR morphologies (see Figures 1, 2, and 3).  On the basis
of spectroscopic and photometric redshifts, and of redshift estimates
from lensing inversion techniques, we find that almost all 15 $\mu$m
sources are behind the cluster lens. 

We estimate the number counts at 15 $\mu$m, after correcting for
incompleteness (a non-uniform correction over the field, due to the
variable lensing amplification and distortion; see Altieri et al. 1999
and Metcalfe et al. 1999). Our number counts are in good agreement
with those derived in empty fields (the Lockman hole, Elbaz et
al. 1999, and the Hubble Deep Field, Aussel et al. 1999).  Moreover,
the number counts in our deepest field -- reaching deeper than any
other survey -- show no sign of flattening, being close to $-1.5 \pm
0.3$ down to 35 $\mu$Jy. Fitting such a steep slope requires strong
evolution models. 

Integrating the 15 $\mu$m number counts over the whole flux range
(0.03--50 mJy) covered by ISOCAM surveys (including ours), we estimate
the resolved background MIR light at 15 $\mu$m: $3.3 \pm 1.3 \times
10^{-9}$ W m$^{-2}$ sr$^{-1}$. This value is very close to the upper
limit set by the gamma-CMBR photon-photon pair production (Stanev \&
Franceschini 1998), and consistent with the predictions from the model
of Tan et al. (1999).

\begin{figure*}
\includegraphics{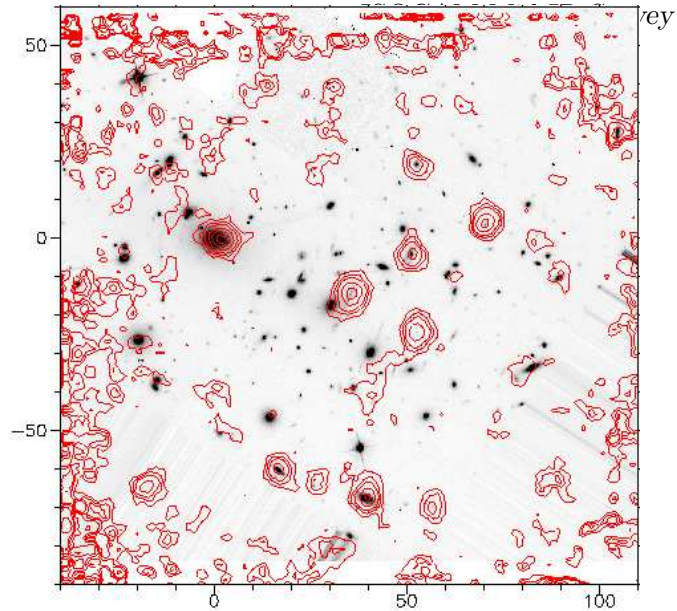}
\vspace{7cm}
\caption{ISOCAM 15 $\mu$m contours over an HST (F814W) I-band image of
A2390. Note the ISOCAM detection of a knot of the straight arc near the
center of the field. Units on both axes
are in arcsec. From Altieri et al. (1999).}
\end{figure*}

\section{Conclusions}
We have detected a population of strong MIR emitters at high redshifts
which cannot be fitted to the local IRAS counts with no evolution
models.  The nature of these faint MIR sources is still unclear. They
could be dust-enshrouded AGN's or dust-enshrouded starbursts, or both.
Recently, Roche \& Eales (1999) and Tan et al. (1999) have tried
fitting the IR counts by invoking a population of starburst galaxies
at high $z$ created by the numerous galaxy mergings predicted in a
hierarchical clustering scenario.  In order to elucidate this issue,
we have recently performed high spatial resolution observations with
ISAAC at the VLT for two of these sources, and submitted an XMM
proposal for distinguishing the AGN's by their strong X-ray
emission. Observations have also been performed at the CFHT to help
determine K-band morphology for several sources.

Counts of MIR sources allow us to estimate a MIR background which is
slightly less than 50~\% of the I-band background.  Extrapolating the
MIR background to the far-IR, assuming typical galaxy spectral energy
distributions, leads to the conclusion that the total cosmic IR
background is actually larger than the optical background.  Dust
processing of stellar radiation is therefore much more important in
distant galaxies than locally.

MIR surveys are therefore essential for tracing the evolution of
galaxies at high redshifts. Great caution must be taken when trying to
infer the global star formation history of the Universe from UV
luminosities, as these must be seriously affected by extinction.

\end{document}